\documentclass[aps,prd,showpacs,amssymb]{revtex4}
\usepackage{graphicx,color}
\usepackage{dcolumn}
\usepackage{bm}

\begin{document}

\draft

\title{ Lepton Flavor Violating Decays  as Probes of Neutrino Mass Spectra and Heavy Majorana Neutrino Masses}

\author{\footnotesize Wan-lei Guo\footnote{Email address: guowl@itp.ac.cn}}

\affiliation{  Institute of Theoretical Physics, Chinese Academy of
Sciences,  Beijing 100080, China }

\begin{abstract}
We investigate the lepton flavor violating (LFV) rare decays in the
supersymmetric minimal seesaw model in which the
Frampton-Glashow-Yanagida ansatz is incorporated. The branching
ratio of $\mu \rightarrow e \gamma$ is calculated in terms of the
Snowmass Points and Slopes (SPS). We find that the inverted mass
hierarchy is disfavored by all SPS points. In addition, once the
ratio of ${\rm BR}(\tau \rightarrow \mu \gamma)$ to ${\rm BR}(\mu
\rightarrow e \gamma)$ is measured, one may distinguish the normal
and inverted neutrino mass hierarchies, and confirm the masses of
heavy right-handed Majorana neutrinos by means of the LFV processes
and the thermal leptogenesis mechanism. It is worthwhile to stress
that this conclusion is independent of the supersymmetric
parameters.

\end{abstract}


\maketitle

\section{Introduction}

Recent solar \cite{SNO}, atmospheric\cite{SK}, reactor\cite{KM} and
accelerator\cite{K2K} neutrino oscillation experiments have provided
us with very robust evidence that neutrinos are massive and lepton
flavors are mixed. The canonical seesaw mechanism \cite{SEESAW}
gives a very simple and appealing explanation of the smallness of
left-handed neutrino masses -- it is attributed to the largeness of
right-handed neutrino masses. The existence of neutrino oscillations
implies the violation of lepton flavors. Hence the lepton flavor
violating (LFV) decays in the charged-lepton sector, such as $\mu
\to e + \gamma$, should also take place. They are unobservable in
the Standard Model (SM), because their decay amplitudes are expected
to be highly suppressed by the ratios of neutrino masses ($m^{}_i
\lesssim 1$ eV) to the $W$-boson mass ($M^{}_W \approx 80$ GeV). In
the supersymmetric extension of the SM, however, the branching
ratios of such rare processes can be enormously enlarged. Current
experimental bounds on the LFV decays $\mu \to e + \gamma$, $\tau
\to e + \gamma$ and $\tau \to \mu + \gamma$ are\cite{limit1}
\begin{eqnarray}
{\rm BR}(\mu \rightarrow e \gamma) \; & < & \; 1.2 \times 10^{-11}
\; ,
\nonumber \\
{\rm BR}(\tau \rightarrow e \gamma) \; & < & \;  1.1 \times 10^{-7}
\; ,
\nonumber \\
{\rm BR}(\tau \rightarrow \mu \gamma) \; & < & \;  6.8 \times
10^{-8} \; .
\end{eqnarray}
The sensitivities of a few planned experiments\cite{limit2} may
reach ${\rm BR}(\mu \rightarrow e \gamma) \lesssim 1.3 \times
10^{-13}$, ${\rm BR}(\tau \rightarrow e \gamma) \sim {\cal
O}(10^{-8})$ and ${\rm BR}(\tau \rightarrow \mu \gamma) \sim {\cal
O}(10^{-8})$.

For simplicity, here we work in the framework of the minimal
supergravity (mSUGRA) extended with two heavy right-handed Majorana
neutrinos. Then all the soft breaking terms are diagonal at high
energy scales, and the only source of lepton flavor violation in the
charged-lepton sector is the radiative correction to the soft terms
through the neutrino Yukawa couplings. In other words, the
low-energy LFV processes $l^{}_j \to l^{}_i + \gamma$ are induced by
the RGE effects of the slepton mixing. The branching ratios of
$l^{}_j \to l^{}_i + \gamma$ are given by \cite{Casas,LFV}
\begin{equation}
{\rm BR}(l^{}_j \rightarrow l^{}_i \gamma) \approx
\frac{\alpha^3}{G^2_{\rm F} m^8_{\rm S}}  \left[ \frac{3 m^2_0 +
A^2_0}{8 \pi^2 v^2 \sin^2\beta} \right]^2 |C^{}_{ij}|^2 \tan^2 \beta
\; ,
\end{equation}
where $m^{}_0$ and $A^{}_0$ denote the universal scalar soft mass
and the trilinear term at $\Lambda^{}_{\rm GUT}$, respectively. In
addition, $m_{\rm S}$ is a typical mass of superparticles, can be
approximately written as \cite{Petcov}
\begin{eqnarray}
m^8_{\rm S} \; \approx \; 0.5m^2_0 M^2_{1/2} (m^2_0 +
0.6M^2_{1/2})^2 \;
\end{eqnarray}
with $M^{}_{1/2}$ being the gaugino mass; and
\begin{eqnarray}
C^{}_{ij} \; = \; \sum_k (M^{}_{\rm D})^{}_{ik} (M^*_{\rm
D})^{}_{jk} {\rm ln} \frac{\Lambda^{}_{\rm GUT}}{M^{}_k} \;
\end{eqnarray}
with $\Lambda_{\rm GUT} = 2.0 \times 10^{16}$ GeV to be fixed in our
calculations. $M^{}_{\rm D}$ and $M^{}_i$ (for $i = 1,2$) represent
the Dirac neutrino mass matrix and the masses of  the heavy
right-handed Majorana neutrinos, respectively.

To calculate the branching ratios of $l^{}_j \to l^{}_i + \gamma$,
we need to know the following parameters in the framework of the
mSUGRA: $M^{}_{1/2}$, $m^{}_0$, $A^{}_0$, $\tan \beta$ and ${\rm
sign } (\mu)$. These parameters can be constrained from cosmology
(by demanding that the proper supersymmetric particles should give
rise to an acceptable dark matter density) and low-energy
measurements (such as the process $b \rightarrow s + \gamma$ and the
anomalous magnetic moment of muon $g_\mu -2$). Here we adopt the
Snowmass Points and Slopes  (SPS) \cite{SPS} listed in Table. 1.
These points and slopes are a set of benchmark points and parameter
lines in the mSUGRA parameter space corresponding to different
scenarios in the search for supersymmetry at present and future
experiments.

Since the canonical seesaw models are usually pestered with too many
parameters, specific assumptions have to be made for $M^{}_{\rm D}$
or $M^{}_i$ so as to calculate the LFV rare decays \cite{LFV1}. In
addition, the LFV processes have also been discussed in the
supersymmetric minimal seesaw model (MSM)
\cite{Ibarra2,Ibarra1,Raidal,LFV2}, in which only two heavy
right-handed Majorana neutrinos are introduced \cite{FGY,2RHN}. In
the MSM, all model parameters can in principle be fixed by use of
the LFV rare decays and eletric dipole moment of the electron
\cite{Ibarra2}. Ibarra and Ross discuss that the LFV  processes may
constrain the masses of right-handed heavy majorana neutrinos
\cite{Ibarra1}. In the Frampton-Glashow-Yanagida \cite{FGY} (FGY)
ansatz, Raidal and Strumia have given analytic approximations to the
LFV rare decays for the normal hierarchy case \cite{Raidal}.  As
shown in the literature \cite{GUO1}, the FGY ansatz only includes
three unknown parameters: $M^{}_1$, $M^{}_2$ and the smallest mixing
angle $\theta^{}_z$. In this paper, we are going to numerically
compute the LFV processes in terms of $\theta^{}_z$ for both normal
and inverted hierarchies in the FGY ansatz. In the following parts,
we shall show that the branching ratio of $\mu \rightarrow e \gamma$
may distinguish the neutrino mass hierarchies\footnote{We may also
distinguish the normal and inverted neutrino mass hierarchies
through measuring $\theta_z$, the Jarlskog parameter of CP violation
($J_{\rm CP}$) and the effective mass of neutrinoless double beta
decay ($\langle m \rangle_{ee}$)\cite{GUO1}.}. In addition, once the
ratio of ${\rm BR}(\tau \rightarrow \mu \gamma)$ to ${\rm BR}(\mu
\rightarrow e \gamma)$ is measured, one may distinguish the normal
and inverted neutrino mass hierarchies, and confirm the masses of
two heavy right-handed Majorana neutrinos by means of the LFV
processes and the thermal leptogenesis mechanism\cite{FY}. It is
worthwhile to stress that this conclusion is independent of the
mSUGRA parameters. The remaining parts of this paper are organized
as follows. In Section II, we briefly describe the main features of
the FGY ansatz and the thermal leptogenesis in the minimal
supersymmetric standard model (MSSM). In Section III, the branching
ratio of $\mu \rightarrow e \gamma$ is computed in terms of the SPS.
In Section IV, we numerically calculated the ratio of ${\rm BR}(\tau
\rightarrow \mu \gamma)$ to ${\rm BR}(\mu \rightarrow e \gamma)$.
Finally the summary  are given in Section V.

\section{The FGY Ansatz and the Thermal Leptogenesis}

In the supersymmetric extension of the MSM, two heavy right-handed
Majorana neutrinos $N^{~}_{i \rm R}$ (for $i = 1, 2$) are introduced
as the $SU(2)^{}_{\rm L}$ singlets. The Lagrangian relevant for
lepton masses can be written as \cite{Mei}
\begin{equation}
-{\cal L}^{}_{\rm lepton} \; =\; \overline{l^{~}_{\rm L}} Y^{~}_l
E^{~}_{\rm R} H^{}_1 + \overline{l^{~}_{\rm L}} Y^{~}_\nu N^{~}_{\rm
R} H^{}_2 + \frac{1}{2} \overline{N^{\rm c}_{\rm R}} M^{~}_{\rm R}
N^{~}_{\rm R} + {\rm h.c.} \; ,
\end{equation}
where $l_{\rm L}$ denotes the left-handed lepton doublet, while
$E^{~}_{\rm R}$ and $N^{~}_{\rm R}$ stand respectively for the
right-handed charged-lepton and neutrino singlets. $H^{}_1$ and
$H^{}_2$ (with hypercharges $\pm 1/2$) are the MSSM Higgs doublet
superfields. After the spontaneous gauge symmetry breaking, one
obtains the charged-lepton mass matrix $M^{~}_l = v_1 Y^{~}_l$ and
the Dirac neutrino mass matrix $M^{~}_{\rm D} = v_2 Y^{~}_\nu$. Here
$v^{}_i$ being the vev of the Higgs doublet $H^{}_i$ (for $i=1, 2$).
An important parameter $\beta$ is defined by $\tan \beta \equiv
v^{}_2/v^{}_1$ or $\sin \beta \equiv v_2/v$ with $v \simeq 174 ~
{\rm GeV}$. The heavy right-handed Majorana neutrino mass matrix
$M^{~}_{\rm R}$ is a $ 2 \times 2 $ symmetric matrix and $M^{~}_{\rm
D}$ is a $3 \times 2$ matrix. Without loss of generality, we work in
the flavor basis where $M^{~}_l$ and $M^{~}_{\rm R}$ are both
diagonal, real and positive; i.e., $M^{~}_l = {\rm Diag}\{m^{~}_e,
m^{~}_\mu, m^{~}_\tau \}$ and $M^{~}_{\rm R} = {\rm Diag}\{M^{~}_1,
M^{~}_2\}$. The seesaw relation \cite{SEESAW}
\begin{eqnarray} M^{~}_\nu = - M^{~}_{\rm D} M^{-1}_{\rm
R} M^T_{\rm D}
\end{eqnarray}
remains valid.  Note that this canonical seesaw relation holds up to
the accuracy of ${\cal O}(M^2_{\rm D}/M^2_{\rm R})$ \cite{XZ}. Since
$M^{~}_{\rm R}$ is of rank 2, $M^{}_\nu$ is also a rank-2 matrix
with $|{\rm Det}(M^{~}_\nu)| = m^{~}_1 m^{~}_2 m^{~}_3 = 0$, where
$m^{~}_i$ (for $i=1, 2, 3$) are the masses of three light neutrinos.
As for three neutrino masses $m^{}_i$, the solar neutrino
oscillation data have set $m^{~}_2
> m^{~}_1$ \cite{SNO}. Now that the lightest neutrino in the MSM must
be massless, we are then left with either $m^{~}_1=0$ (normal mass
hierarchy) or $m^{~}_3 =0$ (inverted mass hierarchy). With the
best-fit values  $\Delta m^{2}_{\rm sun} \equiv m_2^2 - m_1^2 =
8.0\times 10^{-5}~{\rm eV}^2$ and $\Delta m^{2}_{\rm atm} \equiv |
m_3^2 - m_2^2 | = 2.5\times 10^{-3}~{\rm eV}^2$ \cite{FIT}, one can
numerically calculate the neutrino masses \cite{GX}.

In the FGY ansatz \cite{FGY},  $M^{}_{\rm D}$ is taken to be of the
form
\begin{equation}
M^{}_{\rm D} \; =\; \left ( \matrix{ a^{}_1 & ~ {\bf 0} \cr a^{}_2 &
~ b^{}_2 \cr {\bf 0} & ~ b^{}_3 \cr } \right ) \; .
\end{equation}
With the help of Eqs.(6) and (7), one may straightforwardly arrive
at
\begin{equation}
M^{}_\nu \; =\; - \left ( \matrix { \displaystyle
\frac{a^2_1}{M^{}_1} & \displaystyle \frac{a^{}_1 a^{}_2}{M^{}_1} &
{\bf 0} \cr \displaystyle \frac{a^{}_1 a^{}_2}{M^{}_1} &
\displaystyle \frac{a^2_2}{M^{}_1} + \frac{b^2_2}{M^{}_2} &
\displaystyle \frac{b^{}_2 b^{}_3}{M^{}_2} \cr {\bf 0} &
\displaystyle \frac{b^{}_2 b^{}_3}{M^{}_2} & \displaystyle
\frac{b^2_3}{M^{}_2} \cr } \right ) \; .
\end{equation}
Without loss of generality, one can always redefine the phases of
left-handed lepton fields to make $a^{}_1$, $b^{}_2$ and $b^{}_3$
real and positive. In this basis, only $a^{}_2$ is complex and its
phase $\phi \equiv \arg(a^{}_2)$ is the sole source of CP violation
in the model under consideration. Because $a^{}_1$, $b^{}_2$ and
$b^{}_3$ of $M^{}_{\rm D}$ have been taken to be real and positive,
$M^{}_\nu$ may be diagonalized in a more general way
\begin{equation}
M^{}_\nu \; =\; \left (P^{}_l V \right ) \left ( \matrix { m^{}_1 &
0 & 0 \cr 0 & m^{}_2 & 0 \cr 0 & 0 & m^{}_3 \cr } \right ) \left
(P^{}_l V \right )^T \; ,
\end{equation}
where $P^{}_l = i {\rm Diag}\{e^{i\alpha}, e^{i\beta},
e^{i\gamma}\}$ and $V$ is lepton flavor mixing matrix \cite{MNS}
parameterized as
\begin{equation} V \; = \; \left ( \matrix{
c^{~}_x c^{~}_z & s^{~}_x c^{~}_z & s^{~}_z \cr - c^{~}_x s^{~}_y
s^{~}_z - s^{~}_x c^{~}_y e^{-i\delta} & - s^{~}_x s^{~}_y s^{~}_z +
c^{~}_x c^{~}_y e^{-i\delta} & s^{~}_y c^{~}_z \cr - c^{~}_x c^{~}_y
s^{~}_z + s^{~}_x s^{~}_y e^{-i\delta} & - s^{~}_x c^{~}_y s^{~}_z -
c^{~}_x s^{~}_y e^{-i\delta} & c^{~}_y c^{~}_z \cr } \right ) \left
( \matrix{ 1 & 0 & 0 \cr 0 & e^{i\sigma} & 0 \cr 0 & 0 & 1 \cr }
\right ) \;
\end{equation}
with $s^{~}_x \equiv \sin \theta^{~}_x$, $c^{~}_x \equiv \cos
\theta^{~}_x$ and so on. $\theta_x \approx \theta_{\rm  sun}$,
$\theta_y \approx \theta_{\rm  atm}$ and $\theta_z \approx
\theta_{\rm  chz}$ hold as a good approximation \cite{XING03}. In
view of the current experimental data, we have $\theta_x = 34^\circ$
and $\theta_y = 45^\circ$ (best-fit values) as well as $\theta_z <
10^\circ$ at the $99 \%$ confidence level \cite{FIT}. It is worth
remarking that there is only a single nontrivial Majorana
CP-violating phase ($\sigma$) in the MSM, as a straightforward
consequence of $m^{}_1 =0$ or $m^{}_3 =0$.

Note that all six phase parameters ($\delta$, $\sigma$, $\phi$,
$\alpha$, $\beta$ and $\gamma$) have been determined in terms of
$r^{}_{23} = m_2/m_3 \approx 0.18$, $\theta^{}_x$, $\theta^{}_y$ and
$\theta^{}_z$ \cite{GUO1}:
\begin{eqnarray}
\delta & = & \pm \arccos \left [ \frac{c^2_y s^2_z - r^2_{23} s^2_x
\left (c^2_x s^2_y + s^2_x c^2_y s^2_z \right )}{2 r^2_{23} s^3_x
c^{}_x s^{}_y c^{}_y s^{}_z} \right ] \; ,
\nonumber \\
\sigma & = & \frac{1}{2} \arctan \left [\frac{c^{}_x s^{}_y
\sin\delta} {s^{}_x c^{}_y s^{}_z + c^{}_x s^{}_y \cos\delta} \right
] \; ,
\nonumber \\
\alpha & = & -\frac{1}{2} \arctan \left [ \frac{r^{2}_{23} s^2_x
c^2_z \sin 2\sigma}{s^2_z + r^2_{23} s^2_x c^2_z \cos 2\sigma}
\right ] \; ,
\nonumber \\
\beta & = & -\gamma - \arctan \left [ \frac{c^{}_x c^{}_y s^{}_z
\sin\delta} {s^{}_x s^{}_y - c^{}_x c^{}_y s^{}_z \cos\delta} \right
] \; ,
\nonumber \\
\gamma & = & \frac{1}{2} \arctan \left [ \frac{s^2_z \sin
2\sigma}{r^2_{23} s^2_x c^2_z + s^2_z \cos 2\sigma} \right ] \; ,
\nonumber \\
\phi & = & \alpha + \beta - \arctan \left [ \frac{s^{}_x c^{}_y
s^{}_z \sin\delta}{c^{}_x s^{}_y + s^{}_x c^{}_y s^{}_z \cos\delta}
\right ]
\end{eqnarray}
for the normal mass hierarchy ($m_1 =0$). Similar results can also
be obtained for the inverted mass hierarchy ($m^{}_3 =0$)
\cite{GUO1}, but we do not elaborate on them here. Because $|\cos
\delta| \leq 1$ must hold, we find  $0.077 \leq s^{}_z \leq 0.086$
($m_1 =0$) and $0.0075 \leq s^{}_z \leq 0.174$ ($m^{}_3 =0$). Hence,
a measurement of the unknown angle $\theta^{}_z$ becomes crucial to
test the model.

In the flavor basis where the mass matrices of charged leptons and
right-handed neutrinos are diagonal, one can calculate the CP
asymmetry in the decays of the lighter right-handed neutrino and
obtain\cite{Covi}
\begin{eqnarray}
\varepsilon^{}_1  & \approx & \frac{3}{8\pi v^2 \sin^2 \beta} \frac{
M^{}_1 |(M^{}_\nu)^{}_{12}|^2 |(M^{}_\nu)^{}_{23}|^2 \sin 2\phi}
{\left \{|(M^{}_\nu)^{}_{11}|^2 + |(M^{}_\nu)^{}_{12}|^2 \right\}
|(M^{}_\nu)^{}_{33}|}
\end{eqnarray}
when $M_1 \ll M_2$,  where we have used
\begin{eqnarray}
a^2_1 & = & M^{}_1 |(M^{}_\nu)^{}_{11}| \; , ~~ |a^{}_2|^2 = M^{}_1
|(M^{}_\nu)^{}_{12}|^2 / |(M^{}_\nu)^{}_{11}| \; ,
\nonumber \\
b^2_3  & = &  M^{}_2 |(M^{}_\nu)^{}_{33}| \; , ~~~ b^2_2 ~ ~ =
M^{}_2 |(M^{}_\nu)^{}_{23}|^2 / |(M^{}_\nu)^{}_{33}| \; .
\end{eqnarray}
Then $\varepsilon^{}_1$ can result in a net lepton number asymmetry
$Y^{}_{\rm  L}$. The lepton number asymmetry $Y^{}_{\rm  L}$ is
eventually converted into a net baryon number asymmetry $Y^{}_{\rm
B}$ via the nonperturbative sphaleron processes\cite{Kuzmin,HT},
which is given by
\begin{eqnarray}
Y^{}_{\rm  B} = \frac{n_{\rm B}- n_{\rm \bar{B}}}{\bf s} = - c\;
Y_{\rm L} = - c \; \frac{\kappa}{g^{}_*} \; \varepsilon^{}_1 ,
\end{eqnarray}
where  $c = 28/79 \approx 0.35$, $g_* =228.75$ is an effective
number characterizing the relativistic degrees of freedom which
contribute to the entropy ${\bf s}$ of the early Universe, and
$\kappa$ accounts for the dilution effects induced by the
lepton-number-violating wash-out processes. The dilution factor
$\kappa$ can be figured out by solving the full Boltzmann
equations\cite{Covi}, however, we take the following approximate
formula\cite{Kappa}:
\begin{equation}
\kappa \; \approx \; 0.3 \left (\frac{10^{-3} ~ {\rm
eV}}{\tilde{m}^{}_1} \right ) \left [ \ln \left (
\frac{\tilde{m}^{}_1}{10^{-3} ~ {\rm eV}} \right ) \right ]^{-0.6}
\end{equation}
with $\tilde{m}^{}_1 = (M^\dagger_{\rm D} M^{}_{\rm D})^{}_{11} /
{M^{}_1}$.  It is clear that $\varepsilon^{}_1$ and $Y^{}_{\rm B}$
only involve two unknown parameters: $M^{}_1$ and
$\theta_z$.\footnote{Note that $\varepsilon^{}_1$ is inversely
proportional to the mSUGRA parameter $\sin^2 \beta$. Because $\tan
\beta \lesssim 3$ is disfavored (as indicated by the Higgs exclusion
bounds\cite{LEP2}), here we focus on $\tan \beta \geq 5$ or
equivalently $\sin^2 \beta \geq 0.96$. Hence $\sin^2 \beta \approx
1$ is a reliable approximation in our discussion.} A generous range
$8.5 \times 10^{-11} \lesssim Y_{\rm B} \lesssim 9.4 \times
10^{-11}$ has been drawn from the recent Wilkinson Microwave
Anisotropy Probe (WMAP) observational data \cite{WMAP}.  It has been
shown that $M_1 > 3.4 \times 10^{10}$ GeV ($m_1 = 0$) and $M_1 > 2.5
\times 10^{13}$ GeV ($m_3 = 0$) are required by the current
observational data of $Y^{}_{\rm B}$ \cite{Mei}. Flavor effects in
the mechanism of thermal leptogenesis have recently attracted a lot
of attention\cite{Flavor}.  Because the $\tau$ Yukawa coupling are
in thermal equilibrium for $10^9 \; {\rm GeV} \lesssim M_1 \lesssim
10^{12}$ GeV, the flavor issue should be taken into account in our
model for the $m_1 = 0$ case. Using the analytic approximate
formulas in the Ref.\cite{Flavor1},  one may calculate the
CP-violating asymmetry $\varepsilon^{N_1}_{\tau \tau}$ and the
corresponding wash-out parameter $K^{N_1}_{\tau \tau}$ for the
$\tau$ lepton doublet in the final states of $N^{}_1$ decays. We
find $\varepsilon^{N_1}_{\tau \tau}=0$ and $K^{N_1}_{\tau \tau} = 0$
because of ${M_{\rm D}}_{31} = 0$,  namely the $N_1$ decays involving
the $\tau$ lepton doublet don't contribute to the baryon asymmetry $Y_{\rm B}$.
On the other hand, the CP-violating asymmetry $\varepsilon^{N_2}_{\tau \tau}$, which is
produced by the out-of-equilibrium decays of $N^{}_2$, vanishes.
Therefore, such flavor effects may be negligible in our paper.

Note that there is in general a potential conflict between achieving
successful thermal leptogenesis and avoiding overproduction of
gravitinos in the MSM with supersymmetry \cite{Gravitino} unless the
gravitinos are heavier than $\sim 10$ TeV \cite{Hamaguchi}. If the
mass scale of gravitinos is of ${\cal O}(1)$ TeV, one must have $M_1
\lesssim 10^8$ GeV. This limit is completely disfavored in the FGY
ansatz. Such a problem could be circumvented in other supersymmetric
breaking mediation scenarios (e.g., gauge mediation \cite{Ibarra1})
or in a class of supersymmetric axion models \cite{AY}, where the
gravitino mass can be much lighter in spite of the very high
reheating temperature. For simplicity, we choose $Y^{}_{\rm B} = 9.0
\times 10^{-11}$ as an input parameter in this paper. Hence, one may
analyze the dependence of $M_1$ on $\theta_z$ from the successful
leptogenesis.

\section{Distinguish the Neutrino Mass Hierarchy}

With the help of Eqs. (4) and (7), $|C^{}_{ij}|^2$ can explicitly be
written as
\begin{eqnarray}
|C^{}_{12}|^2 & = & |a^{}_1|^2 \; |a^{}_2|^2 \; \left (
\displaystyle {\rm ln} \frac{\Lambda^{}_{\rm GUT}}{M^{}_1} \right
)^2 \; ,
\nonumber \\
|C^{}_{13}|^2 & = & 0 \; ,
\nonumber \\
|C^{}_{23}|^2 & = & |b^{}_2|^2 \; |b^{}_3|^2 \; \left (
\displaystyle {\rm ln} \frac{\Lambda^{}_{\rm GUT}}{M^{}_2} \right
)^2  \; .
\end{eqnarray}
Because of  $|C^{}_{13}|^2=0$, we are left with ${\rm BR}(\tau
\rightarrow e \gamma) = 0$. If ${\rm BR}(\tau \rightarrow e \gamma)
\neq 0$ is established from the future experiments, it will be
possible to exclude the FGY ansatz. Using Eq. (13), we reexpress Eq.
(16) as
\begin{eqnarray}
|C^{}_{12}|^2 & = & M_1^2 \; |(M^{}_\nu)^{}_{12}|^2 \; \left (
\displaystyle {\rm ln} \frac{\Lambda^{}_{\rm GUT}}{M^{}_1} \right
)^2 \; ,
\nonumber \\
|C^{}_{23}|^2 & = & M_2^2 \; |(M^{}_\nu)^{}_{23}|^2 \; \left (
\displaystyle {\rm ln} \frac{\Lambda^{}_{\rm GUT}}{M^{}_2} \right
)^2 \; .
\end{eqnarray}
As shown in Section II, $M^{}_1$ may in principle be confirmed by
leptogenesis for given values of $\sin \theta^{}_z$, but $M^{}_2$ is
entirely unrestricted from the successful leptogenesis with $M_2 \gg
M_1$.

We numerically calculate BR($\mu \rightarrow e \gamma$) for
different values of $\sin \theta^{}_z$ by using the SPS points
\footnote{ For the SPS point 5,  Antusch {\it et al.} \cite{LFV1}
have pointed out that the full RGE running results differ by several
orders of magintude from the leading Log approximation results. In
addition, we don't consider a suppression factor of $\sim 15\% $
from the QED correction \cite{Czarnecki}.}. The results are shown in
Fig. 1. Since the SPS points 1a and 1b (or Points 2 and 3) almost
have the same consequence in our scenario, we only focus on Point 1a
(or Point 3). When $\sin \theta^{}_z \rightarrow 0.077$ or $\sin
\theta^{}_z \rightarrow 0.086$, the future experiment is likely to
probe the branching ratio of $\mu \rightarrow e + \gamma$ in the
$m^{}_1 = 0$ case. The reason is that $\sin \theta^{}_z \rightarrow
0.077$ (or $\sin \theta^{}_z \rightarrow 0.086$) implies $\phi
\rightarrow - \pi/2$ (or $\phi \rightarrow 0$). Furthermore, the
successful leptogenesis requires a very large $M^{}_1$ due to
$\varepsilon^{}_1 \propto \sin 2 \phi$. It is clear that all SPS
points are all unable to satisfy ${\rm BR}(\mu \rightarrow e \gamma)
\leq 1.2 \times 10^{-11}$ in the $m^{}_3 = 0$ case. Therefore, we
can exclude the $m^{}_3 =0$ case when the SPS points are taken as
the mSUGRA parameters. When $\sin \theta^{}_z \simeq 0.014$, ${\rm
BR}(\mu \rightarrow e \gamma)$ arrives at its minimal value in the
$m^{}_3 = 0$ case.  For the SPS slopes, larger $M^{}_{1/2}$ yields
smaller ${\rm BR}(\mu \rightarrow e \gamma)$. We plot the numerical
dependence of BR($\mu \rightarrow e \gamma$) on $M^{}_{1/2}$ in Fig.
2, where we have adopted the SPS slope 3 and taken $300 ~ {\rm GeV}
\leq M^{}_{1/2} \leq 1000 ~ {\rm GeV}$. We find that $M^{}_{1/2}
\geq 474 ~ {\rm GeV}$ (or $M^{}_{1/2} \geq 556 ~ {\rm GeV}$) can
result in $ {\rm BR}(\mu \rightarrow e \gamma) \leq 1.2 \times
10^{-11}$ for $\sin \theta^{}_z = 0.014$ (or $\sin \theta^{}_z =
0.1$). For all values of $M^{}_{1/2}$ between 300 GeV and 1000 GeV,
${\rm BR}(\mu \rightarrow e \gamma)$ is larger than the sensitivity
of some planned experiments, which ought to examine the $m^{}_3 =0$
case when the SPS slope 3 is adopted. The same conclusion can be
drawn for the SPS slopes 1a and 2. In view of the present
experimental results on muon $g_\mu - 2$, one may get $M^{}_{1/2}
\lesssim 430$ GeV for $\tan \beta = 10$ and $A^{}_0
=0$\cite{0306219}, implying that the $m^{}_3 =0$ case should be
disfavored.

\section{Determine the Heavy Majorana Neutrino Masses}

In this paper,  an important conclusion is  the masses of two heavy
right-handed Majorana neutrinos ($M_1$ and $M_2$) can be derived
through the leptogenesis and the LFV rare decays. With the help of
Eqs.(2) and (17), one can obtain
\begin{eqnarray}
R = \frac{{\rm BR}(\tau \rightarrow \mu \gamma)}{{\rm BR}(\mu
\rightarrow e \gamma)} = \frac{M_2^2 \; |(M_\nu)_{23}|^2 \; [{\rm
ln} ({\Lambda^{}_{\rm GUT}}/{M^{}_2}) ]^2 }{M_1^2 \;
|(M_\nu)_{12}|^2 \; [{\rm ln} ({\Lambda^{}_{\rm GUT}}/{M^{}_1}) ]^2}
\; .
\end{eqnarray}
Since the successful leptogenesis can be used to fix $M^{}_1$, a
measurement of the above ratio will allow us to determine or
constrain $M^{}_2$. On the other hand, we may calculate the ratio
$R$ by inputting the appropriate $M_2$. It is worthwhile to remark
that the ratio $R$ is independent of the mSUGRA parameters. To
illustration, we show the numerical results of $R$ as functions of
$\sin \theta_z$ in Fig. 3(a) for the $m_1= 0$ case and Fig. 3(b) for
the $m_3= 0$ case, respectively. In the $m_3 = 0$ case, when $\sin
\theta_z \rightarrow 0.0075$,  the phase $\phi \rightarrow 0$ which
implies $M_1$ and $M_2$ may be larger than the GUT scale
$\Lambda^{}_{\rm GUT} = 2.0 \times 10^{16}$ GeV.  So we demand that
$M_1$ and $M_2$ must be less than the GUT scale. Below the scale
$\Lambda^{}_{\rm GUT}$, $M_2^2 ({\rm ln} (\Lambda^{}_{\rm
GUT}/{M^{}_2}))^2 $ obtain its maximum value at
\begin{eqnarray}
M_2 = \Lambda^{}_{\rm GUT} / e = 7.4 \times 10^{15} {\rm GeV} \; .
\end{eqnarray}
Consequently,  we can derive the maximum  value of $R$ with $M_2 =
7.4 \times 10^{15}$ GeV. In Fig. 3(b), the numerical results at the
left side of the red (green) dash-dot line  are ``nonphysical",
because these results are corresponding to $M_2 > 7.4 \times
10^{15}$ GeV for the $M_2 = 10 M_1$  ($M_2 = 50 M_1$) case. We can
obtain the ratio $M_2/M_1$ once $R$ is measured, furthermore
calculate the heavy neutrino mass $M_2$. From Fig. 3,  one may
derive $R < 2 \times 10^9$ for the  $m_1=0$ case and $R < 8 \times
10^3$ for the $m_3=0$ case, respectively. If the future experiments
prove $R \geq 8 \times 10^3$, the inverted mass hierarchy case can
be excluded. It is worthwhile to stress that the above conclusions
are independent of the mSUGRA parameters.

Finally, let us comment on the $M_1 \gg M_2$ case \cite{Raidal}. In
this case, the CP-violating asymmetry $\varepsilon^{}_2$, which is
produced by the out-of-equilibrium decay of $N^{}_2$, may finally
survive. In order to producing the positive cosmological baryon
number asymmetry ($Y_{\rm B} > 0$), $\delta$ of Eq.(11) must take -
sign. The flavor effects may be neglected since the successful
thermal leptogenesis requires $M_2 \geq 10^{12}$ GeV ($m_1 =0$) and
$M_2 \geq 4.4 \times 10^{13}$ GeV ($m_3 =0$). For $M_1 \geq 5 M_2$,
we calculate the branching ratio of $\mu \rightarrow e \gamma$, find
the inverted hierarchy case is disfavored for all SPS points and
slopes. For the normal hierarchy case, on may distinguish the $M_1
\ll M_2$ case ($R \geq 14$) and the $M_1 \gg M_2$ case ($R \lesssim
24$) in terms of the ratio $R$. In addition,  the masses of heavy
neutrinos may be quasi-degenerate \cite{Degenerate}. In the $M_1
\simeq M_2$ case, the so-called resonant leptogenesis may occur
\cite{Resonant}. Such a scenario could allow us to relax the lower
bound on the lighter right-handed Majorana neutrino mass. Therefore,
the branching ratios of $\mu \rightarrow e \gamma$ and $\tau
\rightarrow \mu \gamma$  may be much less than the sensitivities of
a few planned experiments. $R \simeq |(M_\nu)_{23}|^2/
|(M_\nu)_{12}|^2$ can be derived from Eq.(18). Furthermore, one may
obtain $0.04 \lesssim R^{-1} \lesssim 0.07$ (normal hierarchy) and
$0.00045 \lesssim R^{-1} \lesssim 0.22$ (inverted hierarchy).

\section{Summary}

We have analyzed the LFV rare decays in the supersymmetric version
of the MSM in which the FGY ansatz is incorporated. In this
scenario, there are only three unknown parameters $\theta_z$, $M_1$
and $M_2$. The successful leptogenesis may fix $M_1$ for  given
values of $\theta_z$. We have numerically calculated the branching
ratio of $\mu \rightarrow e \gamma$ for different values of $\sin
\theta_z$ by using the SPS Points. Then one can find the inverted
mass hierarchy case is disfavored by all SPS points. For the SPS
slopes, the planned experiments may measure $\mu \rightarrow e
\gamma$ in the $m_3 = 0$ case. In addition, once the ratio of ${\rm
BR}(\tau \rightarrow \mu \gamma)$ to ${\rm BR}(\mu \rightarrow e
\gamma)$ is measured, we may also distinguish the neutrino mass
hierarchies and fix the masses of two heavy right-handed Majorana
neutrinos by means of the LFV processes and the leptogenesis
mechanism. It is worthwhile to stress that this conclusion is
independent of the mSUGRA parameters.

\acknowledgments{ I am indebted to Z.Z. Xing and S. Zhou for
stimulating discussions and reading the manuscript. I am also
grateful to Y.L. Wu  for helpful communication. This work was
supported in part by the National Nature Science Foundation of China
(NSFC) under the grant 10475105 and 10491306.}

\newpage

\begin{table}
\begin{center}
\begin{tabular}{|c|c|c|c|c|l|}
 \hline Point & $M_{1/2}$ & $m_0$ & $A_0$  & $\tan \beta$ &
 Slope \\\hline
 1\,a & 250 & 100 & -100 & 10 &  $m_0 = - A_0 = 0.4 M_{1/2}$, $M_{1/2}$ varies  \\
 1\,b & 400 & 200 & 0 & 30 &    \\
 2 &  300 & 1450 & 0 & 10 &  $m_0 = 2 M_{1/2} + 850 $ GeV, $M_{1/2}$ varies \\
 3 &  400 & 90 & 0 & 10 &    $m_0 = 0.25 M_{1/2} - 10 $ GeV, $M_{1/2}$ varies \\
 4 &  300 & 400 & 0 & 50 &    \\
 5 &  300 & 150 & -1000 & 5 &   \\\hline
\end{tabular}
\end{center} \caption{Some parameters for the Snowmass Points and Slopes (SPS) in the mSUGRA.
The masses are given in GeV.  $\mu$ appeared in the Higgs mass term
has been taken as $\mu
> 0$ for all SPS.}
\end{table}

\newpage

\begin{figure}[tbp]
\begin{center}
\includegraphics[width=10cm,height=10cm,angle=0]{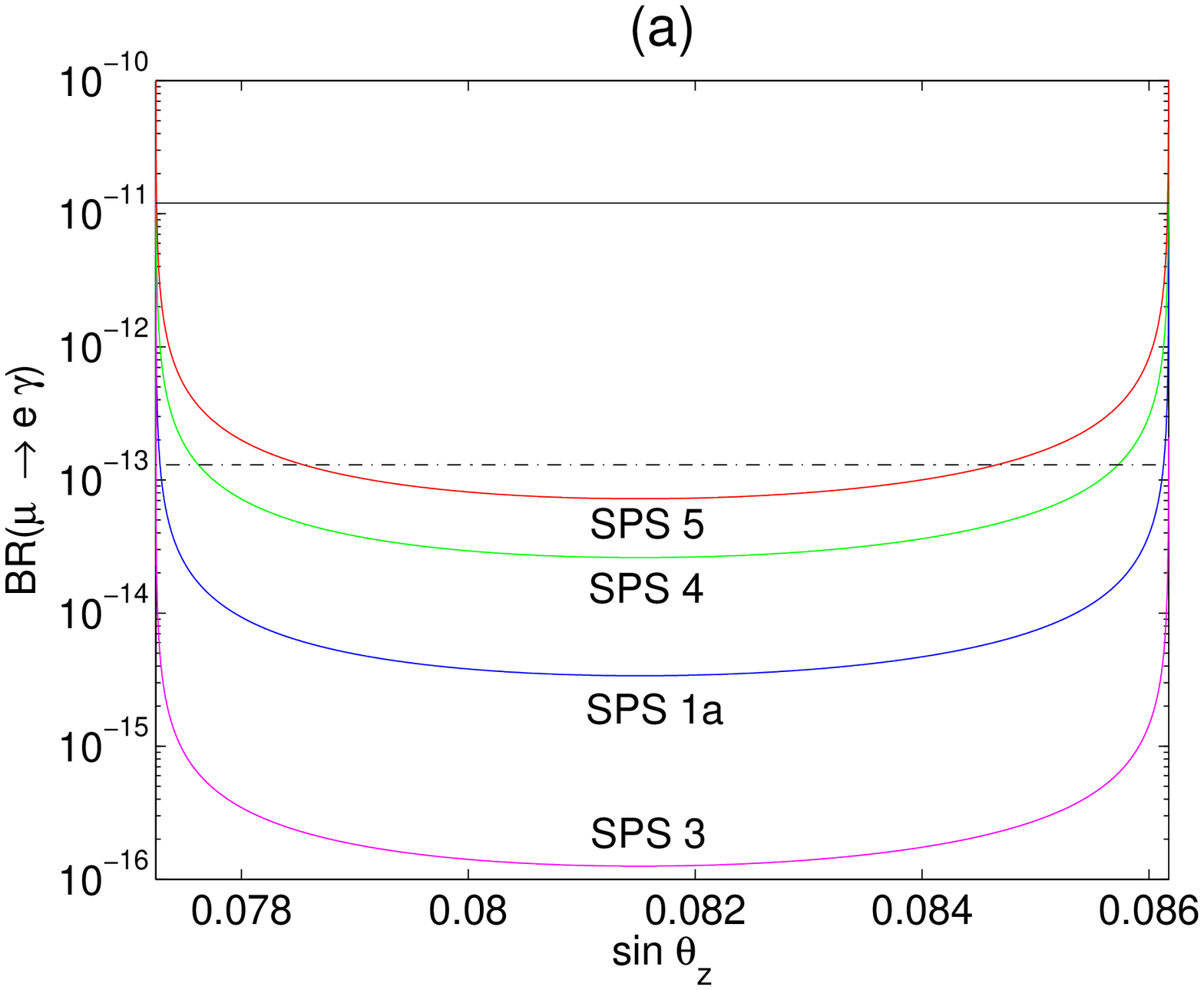}
\includegraphics[width=10cm,height=10cm,angle=0]{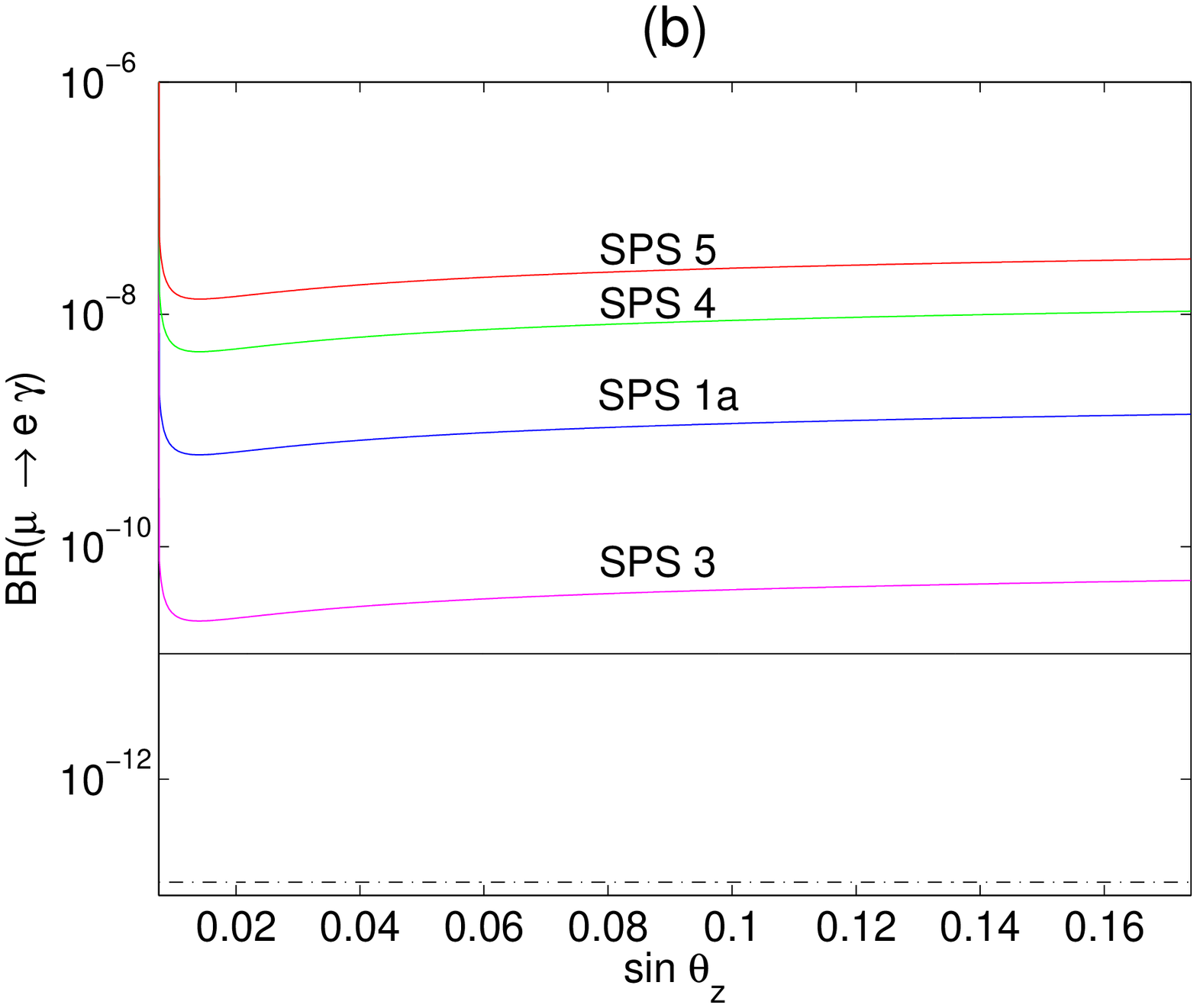}
\end{center}
\caption{Numerical illustration of the dependence of ${\rm BR}(\mu
\rightarrow e \gamma)$ on $\sin \theta^{}_z$: (a) in the $m^{}_1 =0$
case; and (b) in the $m^{}_3 =0$ case. The black solid line and
black dash-dot line denote the present experimental upper bound on
and the future experimental sensitivity to ${\rm BR}(\mu \rightarrow
e \gamma)$, respectively.}
\end{figure}

\newpage

\begin{figure}[tbp]
\begin{center}
\includegraphics[width=10cm,height=10cm,angle=0]{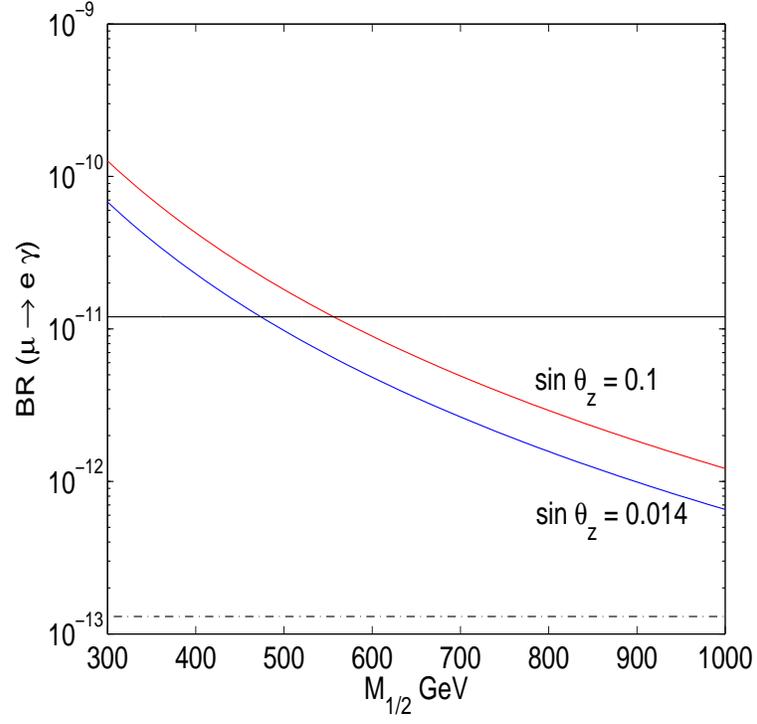}
\end{center}
\caption{ Numerical illustration of the dependence of ${\rm BR}(\mu
\rightarrow e \gamma)$ on $M^{}_{1/2}$ for SPS slope 3 in the
$m^{}_3 =0$ case. The black solid line and black dash-dot line
denote the present experimental upper bound on and the future
experimental sensitivity to ${\rm BR}(\mu \rightarrow e \gamma)$,
respectively.}
\end{figure}

\newpage

\begin{figure}[tbp]
\begin{center}
\includegraphics[width=10cm,height=10cm,angle=0]{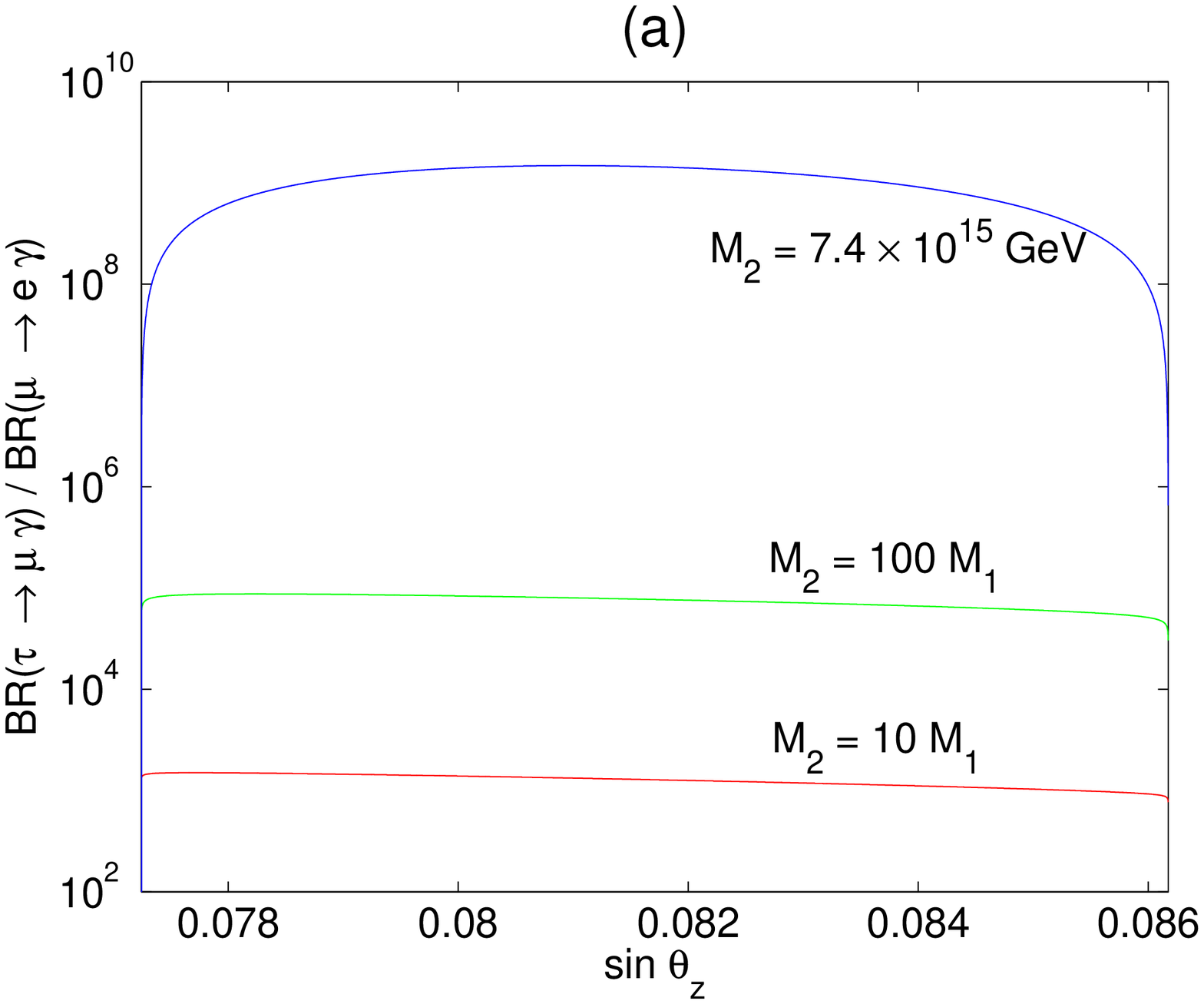}
\includegraphics[width=10cm,height=10cm,angle=0]{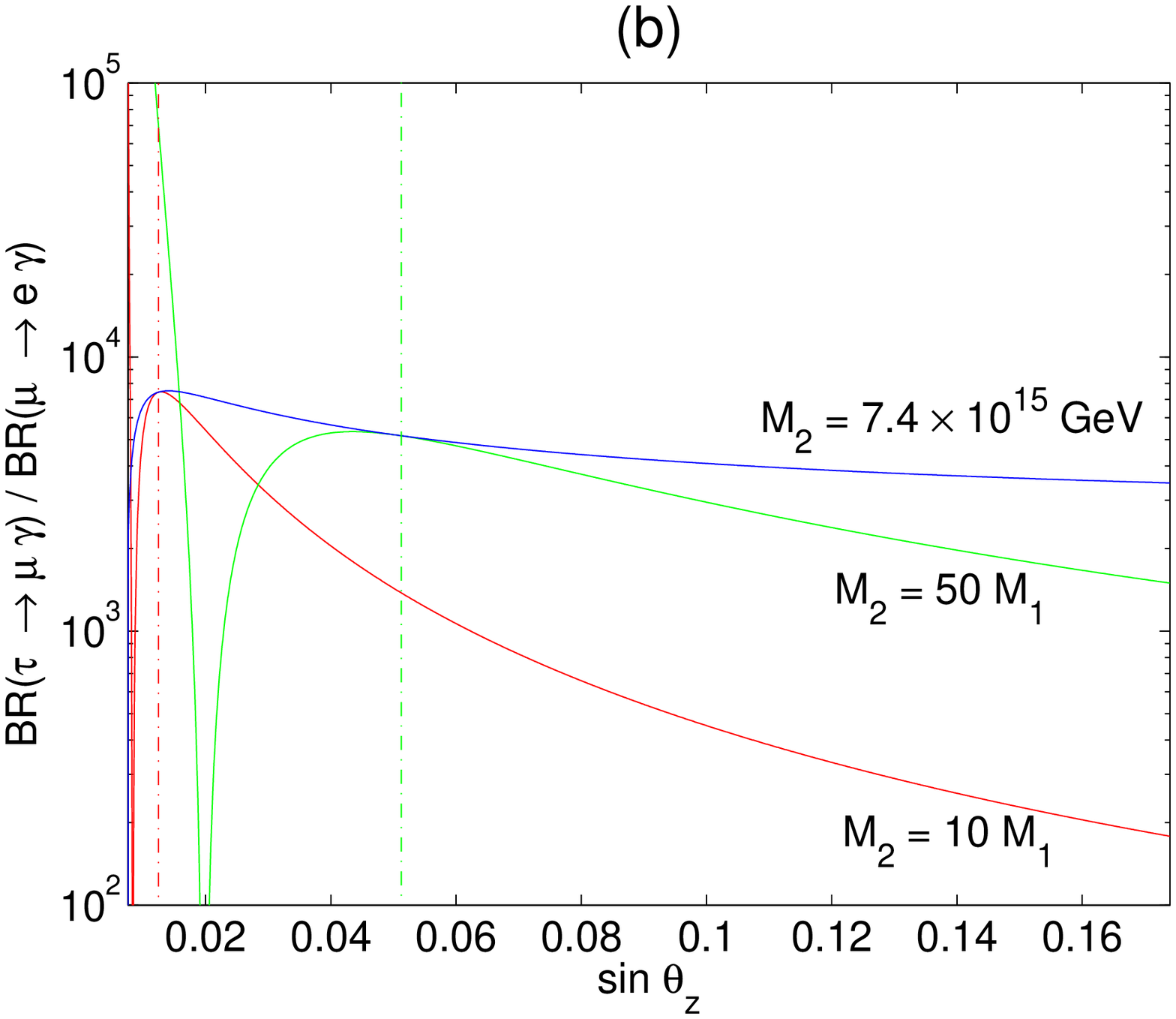}
\end{center}
\caption{Numerical illustration of the dependence of ${\rm BR}(\tau
\rightarrow \mu \gamma)/{\rm BR}(\mu \rightarrow e \gamma)$ on $\sin
\theta^{}_z$: (a) in the $m^{}_1 =0$ case; and (b) in the $m^{}_3
=0$ case.}
\end{figure}

\end{document}